\documentclass[twocolumn,prl,amsmath,amssymb,showpacs,superscriptaddress,floatfix]{revtex4}
\usepackage{graphicx}
\usepackage{bm}
\usepackage{lineno,hyperref}

\usepackage{epsf}
\begin{document}
\title{Stripe phase of two-dimensional core-softened systems: structure recognition}

\author{Yu. D. Fomin, E. N. Tsiok, and V. N. Ryzhov}
\affiliation{ Institute for High Pressure Physics RAS, 108840
Kaluzhskoe shosse, 14, Troitsk, Moscow, Russia}

\date{\today}

\begin{abstract}
In the present paper we discuss a so called stripe phase of
two-dimensional systems which was observed in computer simulation
of core-softened system and in some experiments with colloidal
films. We show that the stripe phase is indeed an oblique crystal
and find out its unit vectors, i.e. we give a full description of
the structure of this crystalline phase.
\end{abstract}


\maketitle


Investigation of two-dimensional (2d) systems is of great interest
for many fundamental and technological issues. They demonstrate
many unusual features which make them especially interesting. The
most striking example is the melting of 2d crystals. While in
three dimensions (3d) melting occurs as the first order phase
transition only, three most plausible mechanisms of melting of 2d
crystals are known at the moment \cite{ufn}. At the same time
until recently it was supposed that 2d systems do not demonstrate
great variety of ordered structures. Only triangular crystal was
observed in experimental studies. However, in the last decade many
experimental investigations demonstrated that 2d and quasi-2d
systems do demonstrate other crystalline structures too. Formation
of square ice when water is confined between two graphite planes
was reported in Ref. \cite{geim}. The square phase of one atom
thick layer of iron on graphite was also observed in Ref.
\cite{iron}. Even more complex phases were observed in 2d
colloidal systems \cite{dobnikar}. However, until now experimental
observations of ordered 2d structures except triangular lattice
are rather rare.

At the same time complex 2d structures are widely discussed in
frames of computer simulation studies. There are numerous
publications which demonstrate existence of different crystalline
and quasicrystalline structures in 2d systems (see, for instance,
\cite{hzmiller,hzch,hzwe,we1,we2,we3,we4,we5,trusket2,trusket3,trusket4,qc1,qc2,qc3}
and references therein). A model core-softened system (smooth
repulsive shoulder SRS) was proposed in Ref. \cite{we-init}. The
potential of this system is defined as

\begin{equation}
  U(r)/\varepsilon= \left( \frac{d}{r} \right)^n+0.5 \left( 1 - tanh(k(r- \sigma))
  \right),
\end{equation}
where $n=14$, $k=10$ and the parameter $\sigma$ determines the
width of the repulsive shoulder of the potential. It was shown
that the phase diagram of three dimensional SRS system is very
complex, with numerous crystalline phases, maxima on the melting
line, etc. The behavior of the same system in 2d was investigated
in Refs. \cite{we1,we2,we3,we4,we5}. According to these
publications, while at $\sigma=1.15$ the system demonstrates only
triangular crystalline phase, at $\sigma=1.35$ square crystal and
dodecagonal quasicrystal are found.

The SRS system was generalized in Ref. \cite{we-attract} by adding
an attractive well to the potential (SRS - attractive well system
- SRS-AW):

\begin{equation}\label{tanh2}
  U(r)/\varepsilon= \left( \frac{d}{r} \right)^n+ \sum_{i=1}^2 \left( 1 - tanh(k_i(r- \sigma_i))
  \right).
\end{equation}
Phase diagrams and anomalous behavior of 3d SRS-AW system with
different parameters of the potential were investigated
\cite{we-attract,we-attract1,we-attract2,we-attract3}. Later on in
a number of publications it was realized that varying the
parameters of SRS-AW system different complex crystalline
structures both in 2d and 3d can be obtained. In Ref.
\cite{str2d3d} a method of finding a potential which stabilizes a
particular crystalline structure was proposed and potentials which
stabilize square and honeycomb lattices were obtained. This method
was used to find parametrization of SRS-AW potential to stabilize
different 2d and 3d structures including Kagome lattice,
snub-square tiling, honeycomb lattice in the case of 2d systems
and cubic and diamond structures for the 3d ones
\cite{trusket1,trusket2,trusket3,trusket4}.

A particular structure which was found in some 2d systems is the
so called stripe phase. Stripe phases are known in many different
system including magnetic films, Langmuir monolayers, polymer
films, etc. In the case of atomic systems the formation of stripe
phase is usually attributed to the competition of short-range
attraction between the particles and long-range dipole-dipole
interaction. Basing on this assumption a model was proposed in
Ref. \cite{camp} where the interaction potential consists of
Lennard-Jones term and long range dipole-dipole term. Rather small
system (500 particles) was studied in this paper. A larger system
of 2000 particles was simulated in addition to check for the
finite size effects. It was discovered that at some
density-temperature points systems of 500 particles form a
lamellar phase. However, in case of larger system the lamellar
phase was not formed (see Fig. 12 of Ref. \cite{camp}). Instead of
this the stripe phase was observed which differs from the lamellar
one in the sense that the threads of the particles becomes curved
rather then linear. Because of this it was concluded that the
lamellar phase was artificially stabilized by the periodic
boundary conditions and the stripe phase is the stable one.
Moreover, it was stated that the stripe phase is a indeed a
cluster fluid. These results were confirmed in the subsequent
publication of the same author \cite{camp1}.

Formation of the stripe phase was discovered in the simplest
core-softened system - repulsive step potential
\cite{malpel,norizoe,norizoe1,dijkstra}. The stripe phase was
observed for several different values of the width of the step. In
particular, in Ref. \cite{malpel} the authors reported a peak of
the heat capacity which appears in the system upon heating. Fig. 3
of Ref. \cite{malpel} reports snapshots of the system above the
peak of the heat capacity and below it. In Fig. 3 of Ref.
\cite{malpel} one can see snapshots of the system below and above
the peak of the heat capacity. The structure below the peak
corresponds to lamellar phase in terms of Ref. \cite{camp} while
the structure above the peak looks like the stripe phase in
\cite{camp}. In this respect the results of these publications
appear to be qualitatively different: while in \cite{camp} the
change from the structure with linear threads of the particles
(lamellar phase) to the structure with curved threads (stripe
phase) is related to the finite size effects, in \cite{malpel}
these structures are separated by the phase transition, i.e. they
are two different phases.



Importantly, most of the authors reported the existence of the
stripe phase but did not try to describe its structure. The only
work where an attempt to describe the structure of the stripe
phase is Ref. \cite{dijkstra} (see eqs. (7) and (8)).

In Ref. \cite{genetic} the ground states of the repulsive step
potential were investigated. The structures were obtained by
genetic algorithms optimization. The authors did not find anything
like stripe phase. However, the values of the step width reported
in \cite{genetic} are 1.5, 3.0, 7.0 and 10.0. In Ref.
\cite{norizoe} it was argued that at these values of the width no
stripe phase is found.

In Ref. \cite{dobnikar} an experimental study of colloidal
particles in magnetic field was performed. The sequence of phases
which the authors observed was very similar to the one obtained in
the computational work \cite{camp}. In particular, it was found
that the experimental system does demonstrate the stripe phase.

In the present paper we perform a computational study of stripe
phase in a core-softened system. We carry out the analysis of its
structure. We find that the stripe phase is in fact a crystalline
structure and find its unit cell.

\begin{table}
\begin{tabular}{|c|c|}
  \hline
  A & 0.01978 \\
  n & 5.49978 \\
  $\lambda_1$ & -0.06066 \\
  $k_1$ & 2.53278 \\
  $d_1$ & 1.94071 \\
  $\lambda_2$ & 1.06271 \\
  $k_2$ & 1.73321 \\
  $d_2$ & 1.04372 \\
  $r_c$ & 3.0 \\
  $P$ & 0.007379 \\
  $Q$ & 0.04986 \\
  $R$ & -0.085054 \\
  \hline
\end{tabular}
\caption{The parameters of the potential \ref{tanh3} used in the
present study.}
\end{table}


Following Ref. \cite{trusket4} let us rewrite the potential
\ref{tanh2} in the following form:

\begin{equation}\label{tanh3}
  U(r)/\varepsilon= A \left( \frac{\sigma}{r} \right)^n+ \sum_{i=1}^2 \lambda_i \left( 1 - tanh(k_i(r/ \sigma- d_i))
  \right) + U_{shift} 
\end{equation}
where $U_{shift}=Pr^2+Qr+R$ is used to make both the potential and
its first and second derivative continuous at the cut-off distance
$r_c$. We use the parameterizations of the potential which
stabilizes the Kagome lattice \cite{trusket4}. The parameters of
the potential are taken from Ref. \cite{trusket4}. For the
convenience of the reader they are given in Table I. We find that
at the densities below the ones where the Kagome lattice is stable
the system demonstrates the stripe phase. The full phase diagram
of this system will be a topic of a subsequent publication.

In the remainder of this paper we use the dimensionless
quantities, which in $2D$ have the form: $\tilde{{\bf r}}\equiv
{\bf r}/\sigma$, $\tilde{P}\equiv P \sigma^{2}/\varepsilon ,$
$\tilde{V}\equiv V/N \sigma^{2}\equiv 1/\tilde{\rho}, \tilde{T}
\equiv k_BT/\varepsilon$, $\tilde{t}= \sigma (m/
\varepsilon)^{1/2}$, where $m=1$ is the mass of the particles
which is used as a unit of mass., etc. In the rest of the article
the tildes will be omitted.

Initially we simulate a system of 5000 particles in a rectangular
box with periodic boundary conditions by means of molecular
dynamics method in order to find the region of stability of the
stripe phase. The system is simulated for $5 \cdot 10^6$ steps.
The time step is set to $dt=0.001$. The first $3 \cdot 10^6$ are
used for equilibration while during the last $2 \cdot 10^6$ we
collect the data. We calculate the equations of state and the
radial distribution functions of the system. The system is
simulated in NVT ensemble (constant number of particles N, volume
V and temperature T). Nose-Hoover thermostat with time parameter
$\tau=0.01$ is used.



\begin{figure}

\includegraphics[width=6cm,height=6cm]{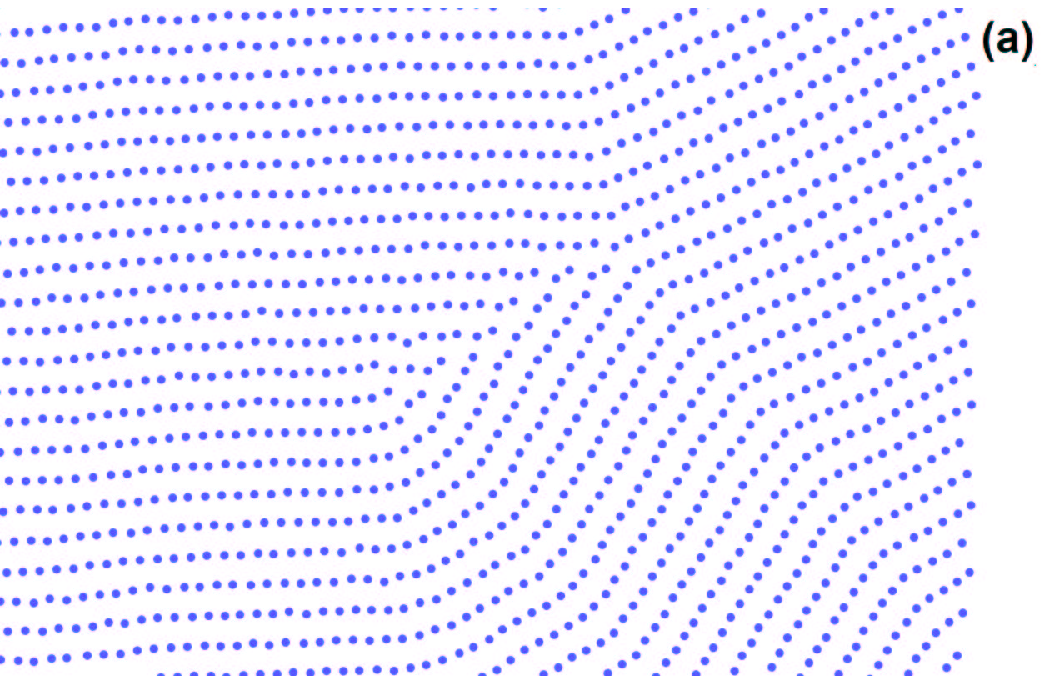}%

\includegraphics[width=6cm,height=6cm]{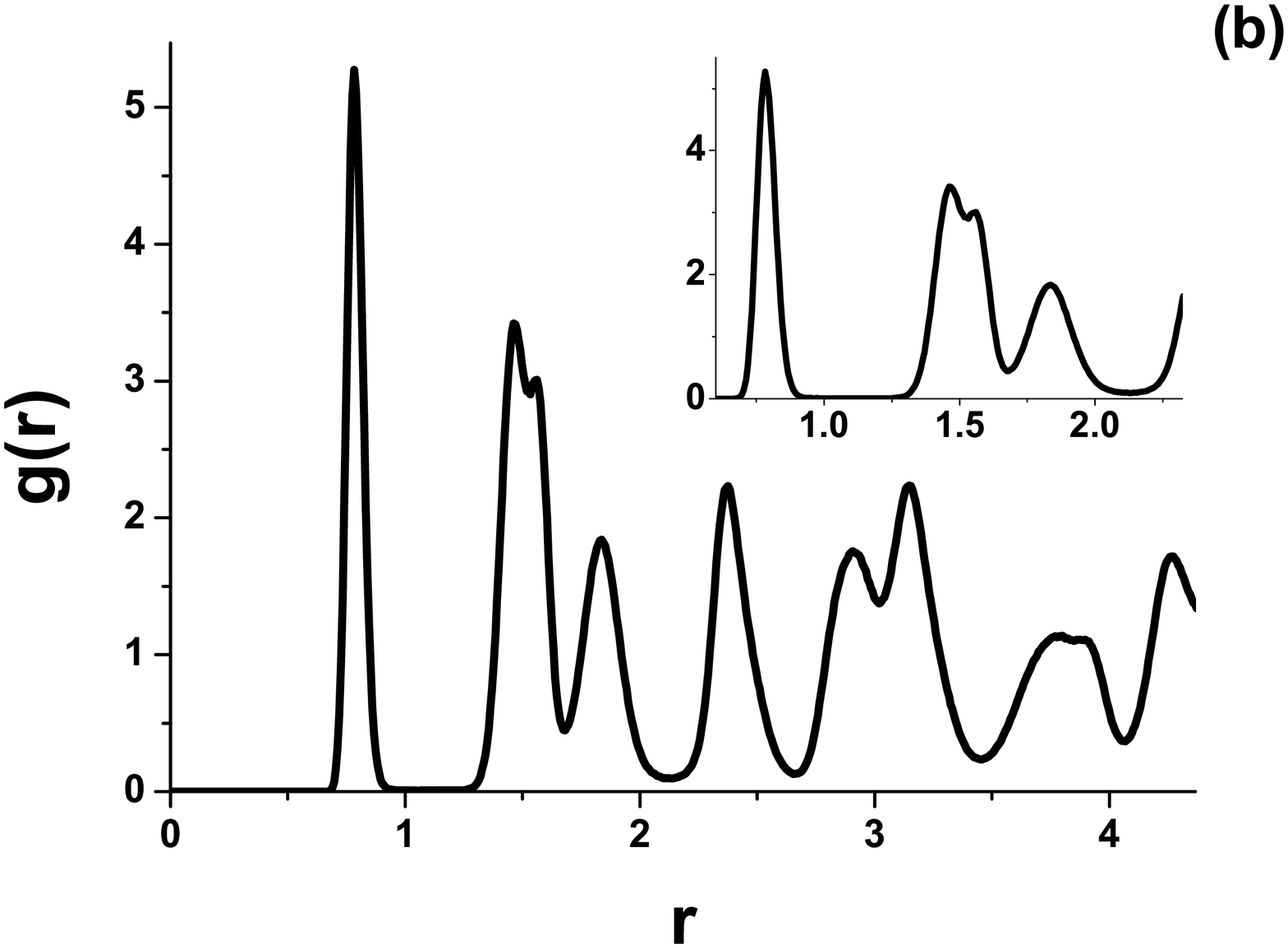}%

\includegraphics[width=6cm,height=4cm]{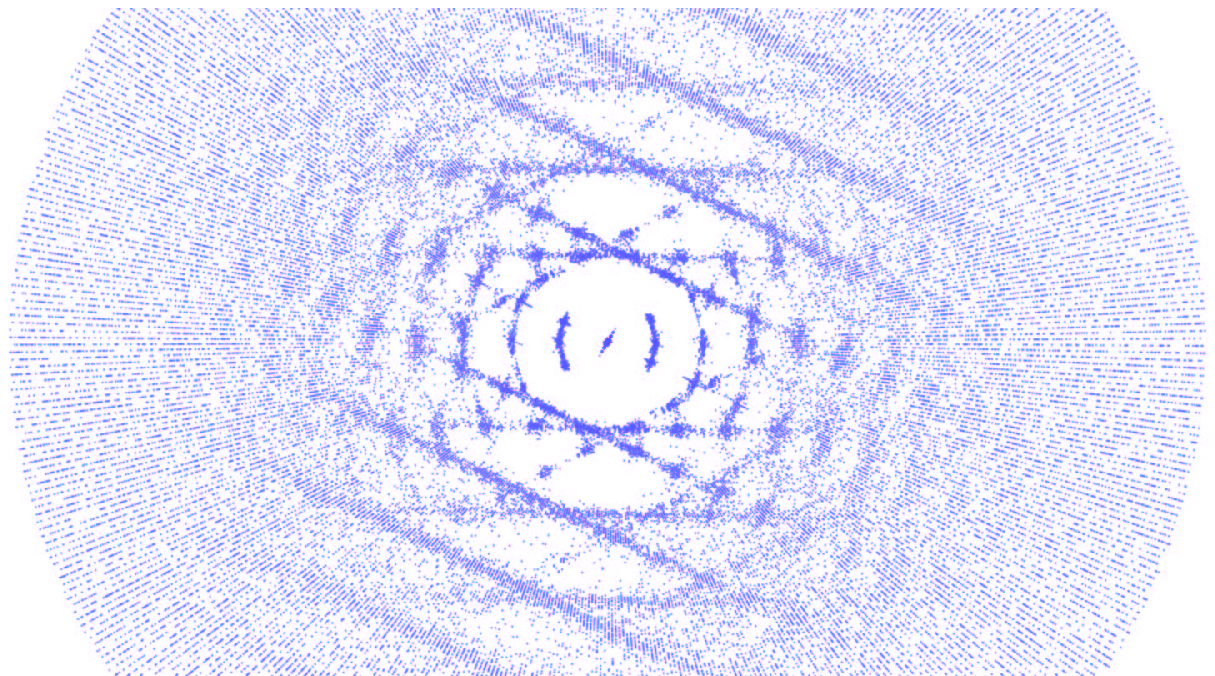}%

\caption{\label{r09t005} (a) A snapshot of the system at
$\rho=0.9$ and $T=0.005$; (b) radial distribution function of the
system at the same parameters. The inset enlarger the region of
small $r$; (c) the diffraction pattern at the same parameters.}
\end{figure}

Fig. \ref{r09t005} shows an instantaneous configuration of the
system at the density $\rho=0.9$ and temperature $T=0.01$. One can
see that this system is in lamellar phase in terminology of Camp
\cite{camp} and stripe phase in terminology of Malescio and
Pelicane \cite{malpel}. Camp observed such a phase in a small
system of 500 particles. In a larger system he found that the
lines of particles deviate from the linear shape and concluded
that lamellar phase is a consequence of the periodic boundary
conditions (see Fig. 12 in Ref. \cite{camp}). We also observe such phase. However, simulating the system for longer time we find that the threads become strainght and the system comes into lamellar phase. Because of this we suppose that the configuration shown in Fig. 12b of Ref. \cite{camp} is not completely equilibrated one.


In Ref. \cite{malpel} it was assumed that such a structure
demonstrates long-range orientational ordering, but not
translational one. Fig. \ref{r09t005} (b) shows the rdf of this
system at the same point $\rho=0.9$ and $T=0.005$. One can see
that it demonstrated clear long-range order. We also calculate the
diffraction pattern of the system. The diffraction pattern is
calculated as $S(\bf{k})= < \frac{1}{N} \left(
\sum_i^Ncos(\bf{kr}_i)\right)^2+\left(
\sum_i^Nsin(\bf{kr}_i)\right)^2>$. The results are shown in Fig.
\ref{r09t005} (c)) and they demonstrate that the system is in
crystalline phase. Therefore, in order to describe this structure
one needs to find the parameters of the unit cell.

\begin{figure}
\includegraphics[width=6cm,height=6cm]{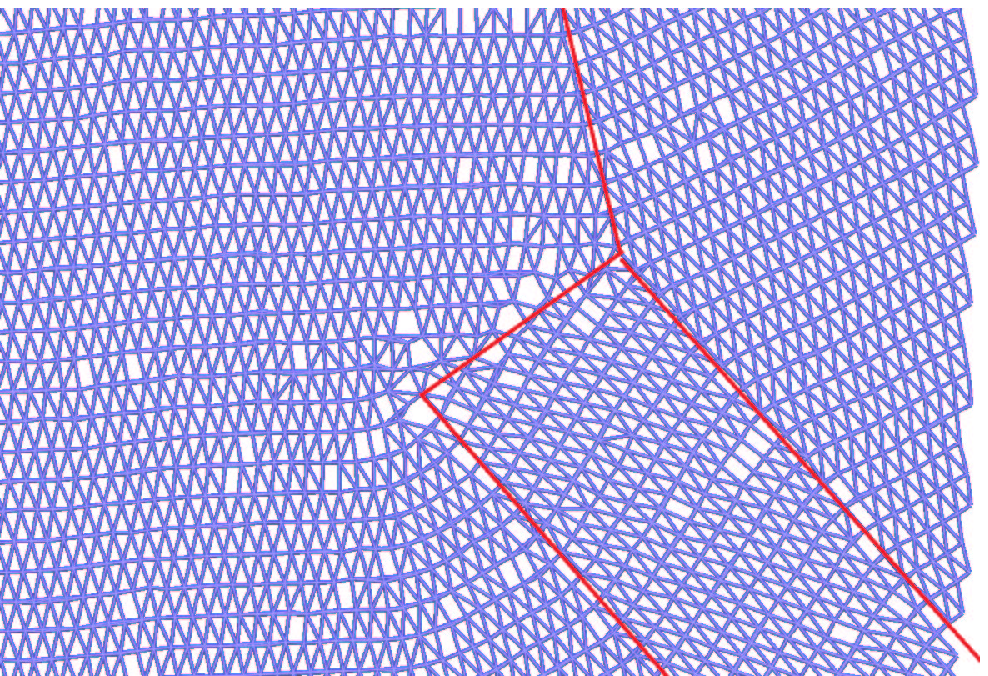}%

\includegraphics[width=6cm,height=6cm]{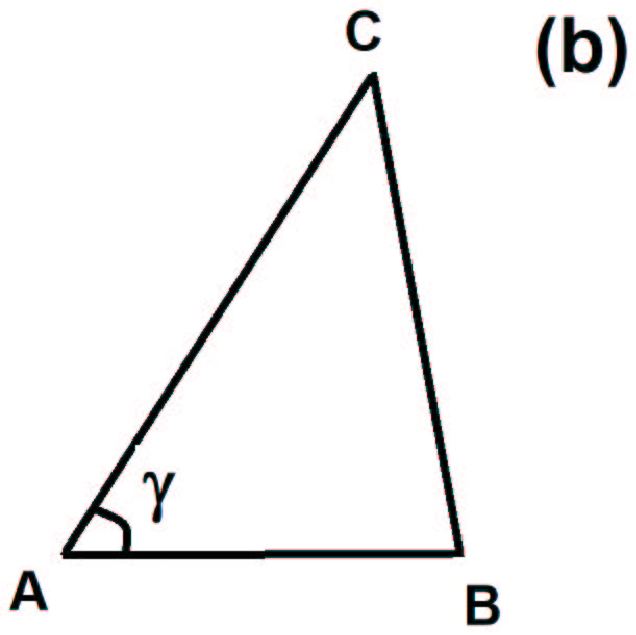}%

\caption{\label{bonds} (a) A snapshot of the system at $\rho=0.9$
and $T=0.005$ drawn by bonds. The lines show the grain boundaries;
(b) a cartoon of the base triangle of the structure. AB=0.78227,
AC=1.56808, BC=1.46189.}
\end{figure}

Let us draw the structure by bonds rather then points (Fig.
~\ref{bonds}). We connect two particles by a bond if they are
closer then some distance $r_b$. The parameter $r_b$ is selected
in such a way that the system demonstrates a clear pattern: below
$r_b$ many particles are not connected or connected to a single
particle only while above it the particles are connected to many
other particles. Here $r_b=1.6$ One can see that the main motif is
a triangle. The sides of the triangle can be taken as the
locations of the first three maxima of the rdf. From Fig.
~\ref{r09t005} we obtain $r_1=0.78227$ (side AB), $r_2=1.46189$
(side BC)  and $r_3=1.56808$ (side AC). From basic geometry we
obtain $cos(\gamma)=0.38$. One can choose the vectors \textbf{AB}
and \textbf{AC} (Fig. ~\ref{bonds} (b)) as the basic vectors.
Their coordinates are $\textbf{a}=\textbf{AB}=(a,0)$, where
$a=AB=0.78227$ and $\textbf{b}=\textbf{AC}=(bcos \gamma,bsin
\gamma$), where $b=AC=1.56808$. Below we will call the structure
with these basic vectors as oblique phase.

We construct the lattice with basis vectors $\bf{a}$ and $\bf{b}$.
It is a rhomboid structure with the tilt angle $\gamma$. In the
present study we use a system which contains 20000 particles. We
simulate the system at constant pressure $P=P_{xx}=P_{yy}$. The
diagonal component of the pressure $P_{xy}$ was set to zero, since
in equilibrium there should not be any tangential stresses in the
system. The sides and the tilt angle could vary independently in
the course of simulation. Parinello-Rahman thermostat is used.
Longer simulations of $5 \cdot 10^7$ steps are performed in order
to ensure that the structure is stable.


From molecular dynamics simulation of the oblique phase we observe that the oblique structure is stable. The limits of stability of this structure at
$T=0.01$ are from $P_{min}=2.3$ and up to $P_{max}=5.0$ . At
larger pressures the system transforms into Kagome lattice, while
at smaller ones into the phase of dimers. The full phase diagram
of this system will be published in the subsequent paper. The
average tilt angle at different pressures varies from $cos(\gamma)
=0.407 $ up to $cos(\gamma) =0.422 $.

\begin{figure}
\includegraphics[width=6cm,height=6cm]{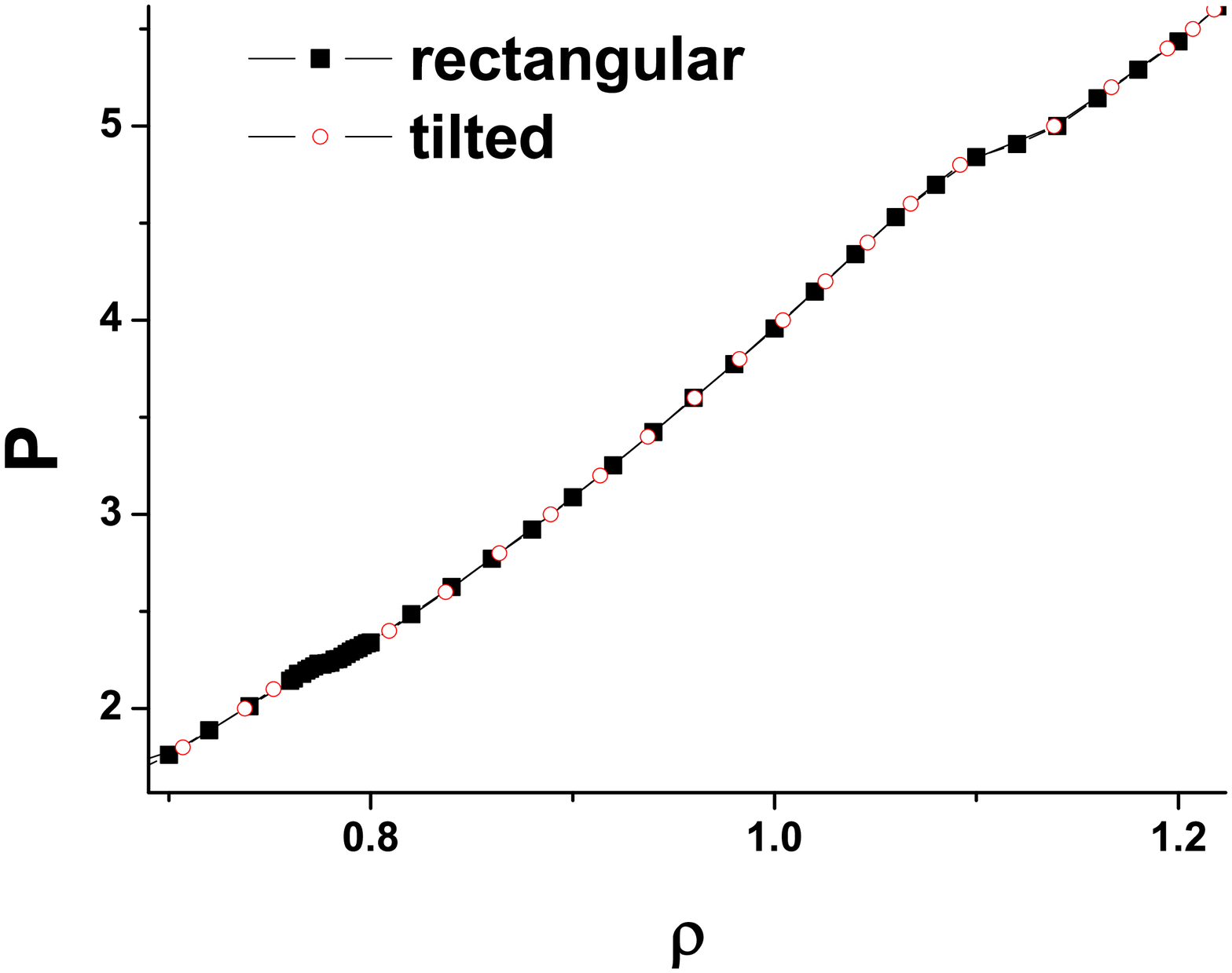}%

\caption{\label{rect} Comparison of equations of states obtained
by simulations in rectangular and tilted boxes at $T=0.01$.}
\end{figure}

Fig. ~\ref{rect} gives a comparison of equations of states
obtained in rectangular and tilted boxes. One can see that these
equations of state are in perfect agreement which means that the
phases obtained in both types of boxes are the same.

\begin{figure}

\includegraphics[width=6cm,height=6cm]{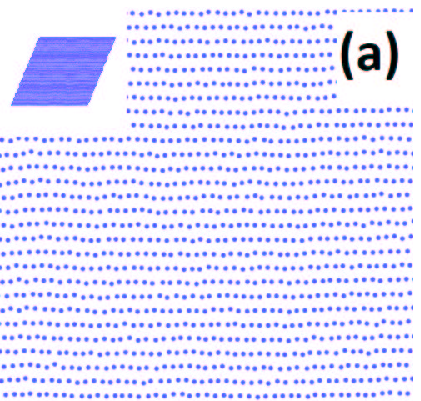}%

\includegraphics[width=6cm,height=6cm]{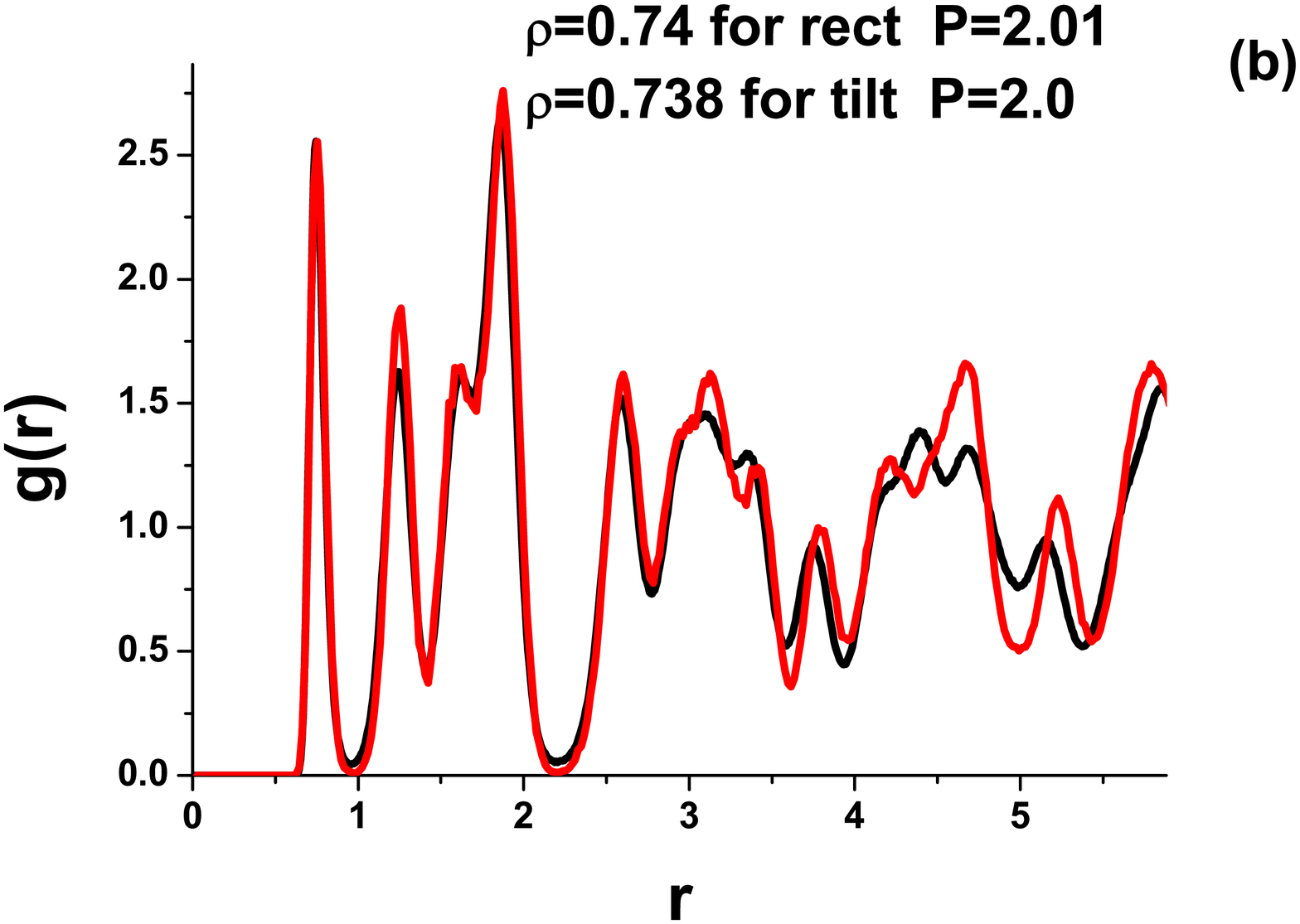}%

\includegraphics[width=6cm,height=6cm]{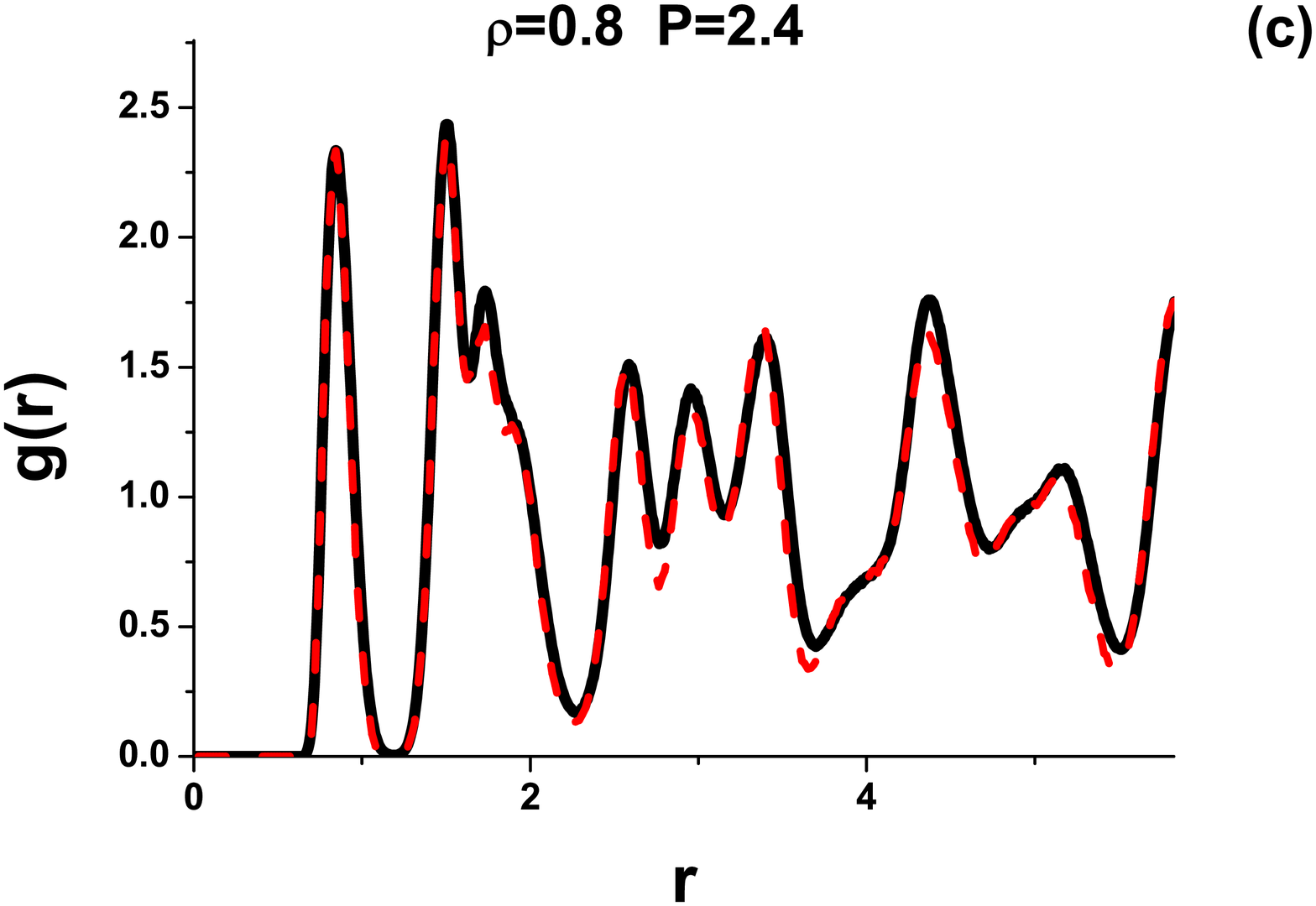}%

\caption{\label{rdf-tilt} (a) Snapshot of the system simulated in
the tilted box at $T=0.01$ and $P=2.4$. In the left upper corner
we show the whole box to demonstrate that it is tilted. The
snapshot itself enlarges a part of the system to see that
particles. (b) Comparison of rdfs of rectangular and tilted
systems at $\rho=0.74$, $T=0.01$; (b) The same at $\rho=0.8$,
$T=0.01$.}
\end{figure}

Fig. ~\ref{rdf-tilt} (a) shows a snapshot of the system simulated
in tilted box at $T=0.01$ and $P=2.4$. This point belongs to the
stripe phase in terminology of Refs. \cite{malpel}. One can see
that the system preserves the crystalline order of the oblique
phase. Figs. ~\ref{rdf-tilt} (b) and (c) compare the rdfs of the
system at two points of stability of the stripe phase computed in
rectangular and tilted boxes. The rdfs of the rectangular and
tilted system are in perfect agreement which confirms that the
stripe phase is a polycrystalline sample of the oblique phase.

\begin{figure}
\includegraphics[width=6cm,height=6cm]{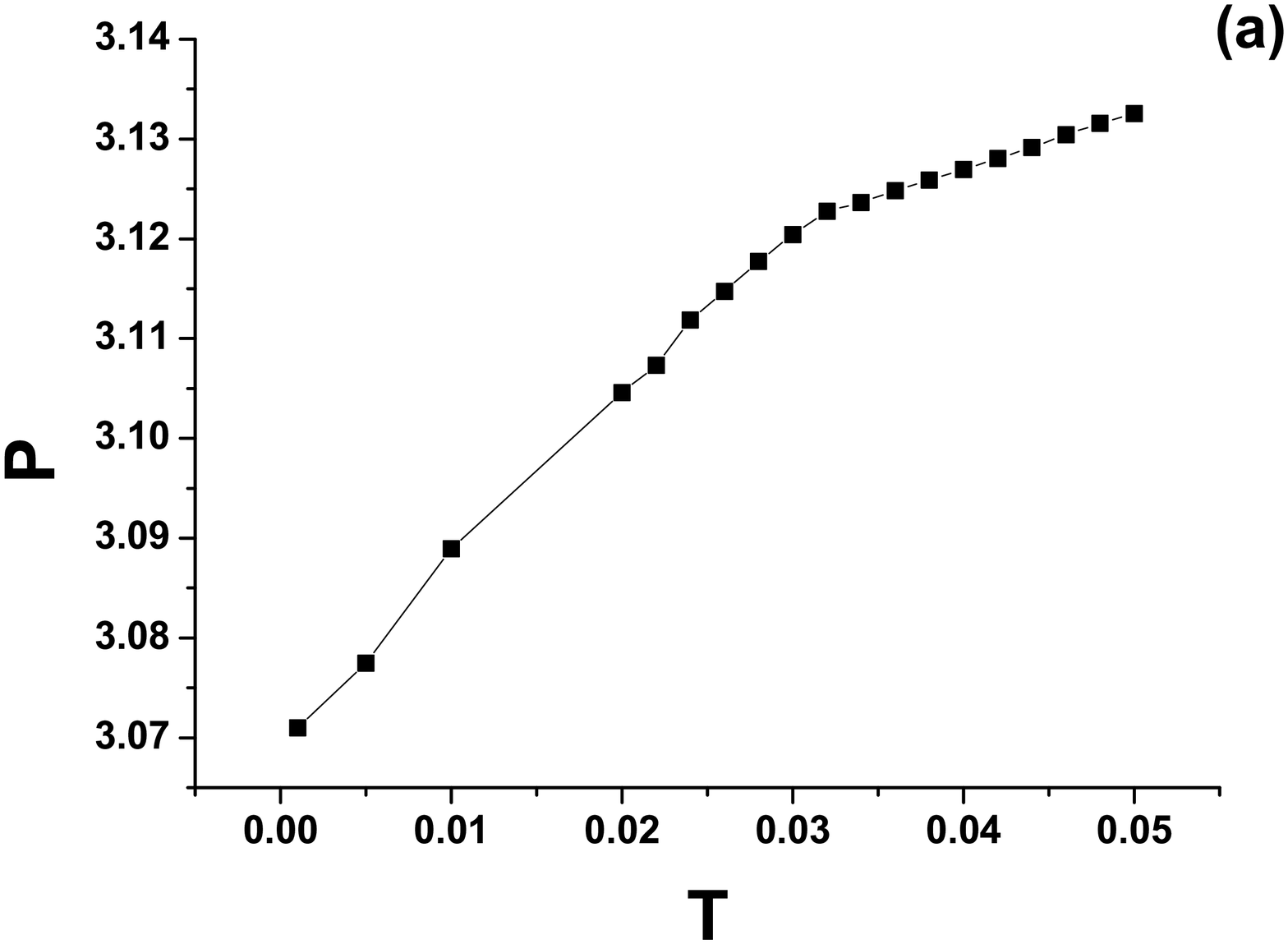}%

\includegraphics[width=6cm,height=6cm]{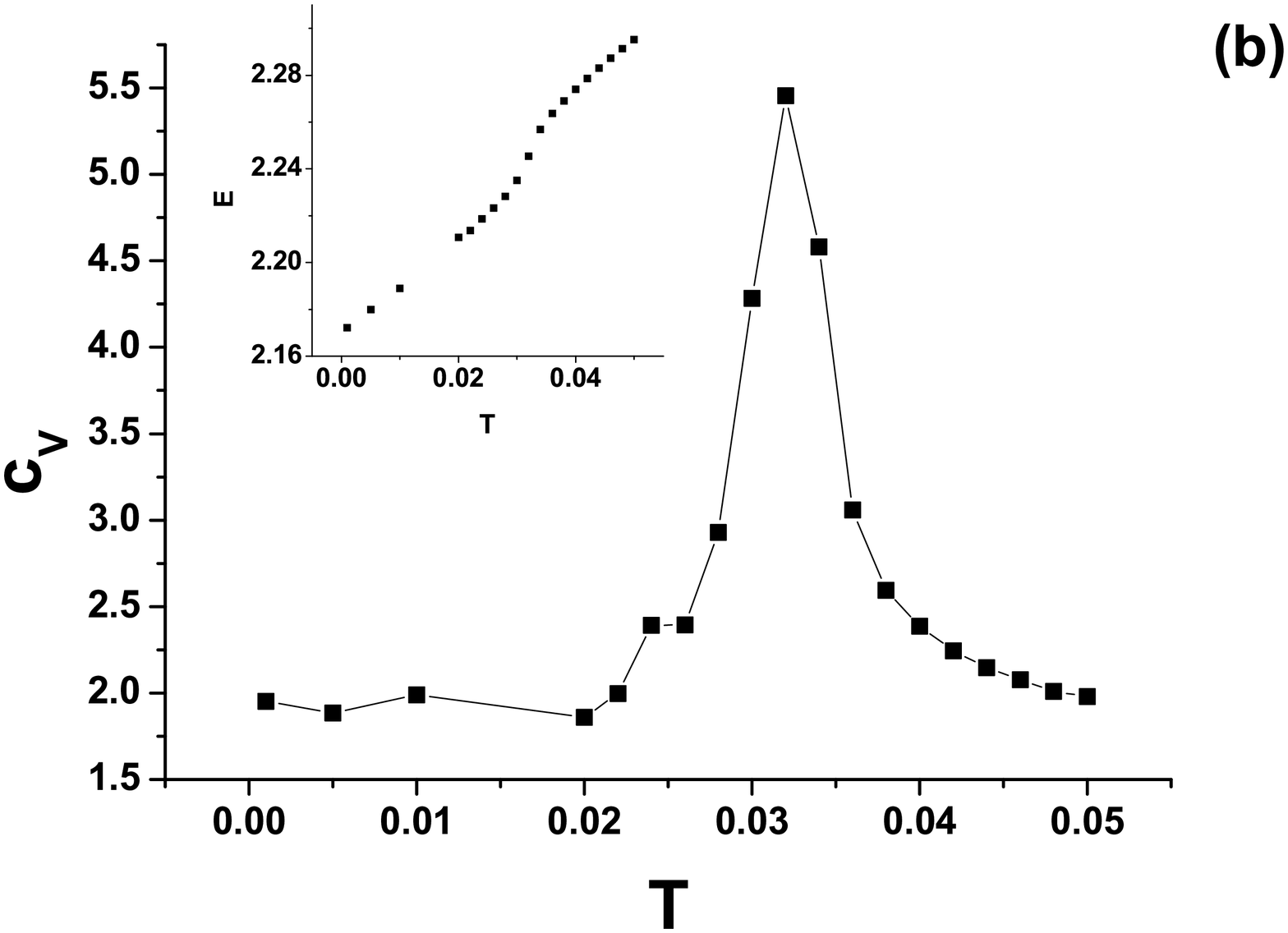}%

\includegraphics[width=6cm,height=6cm]{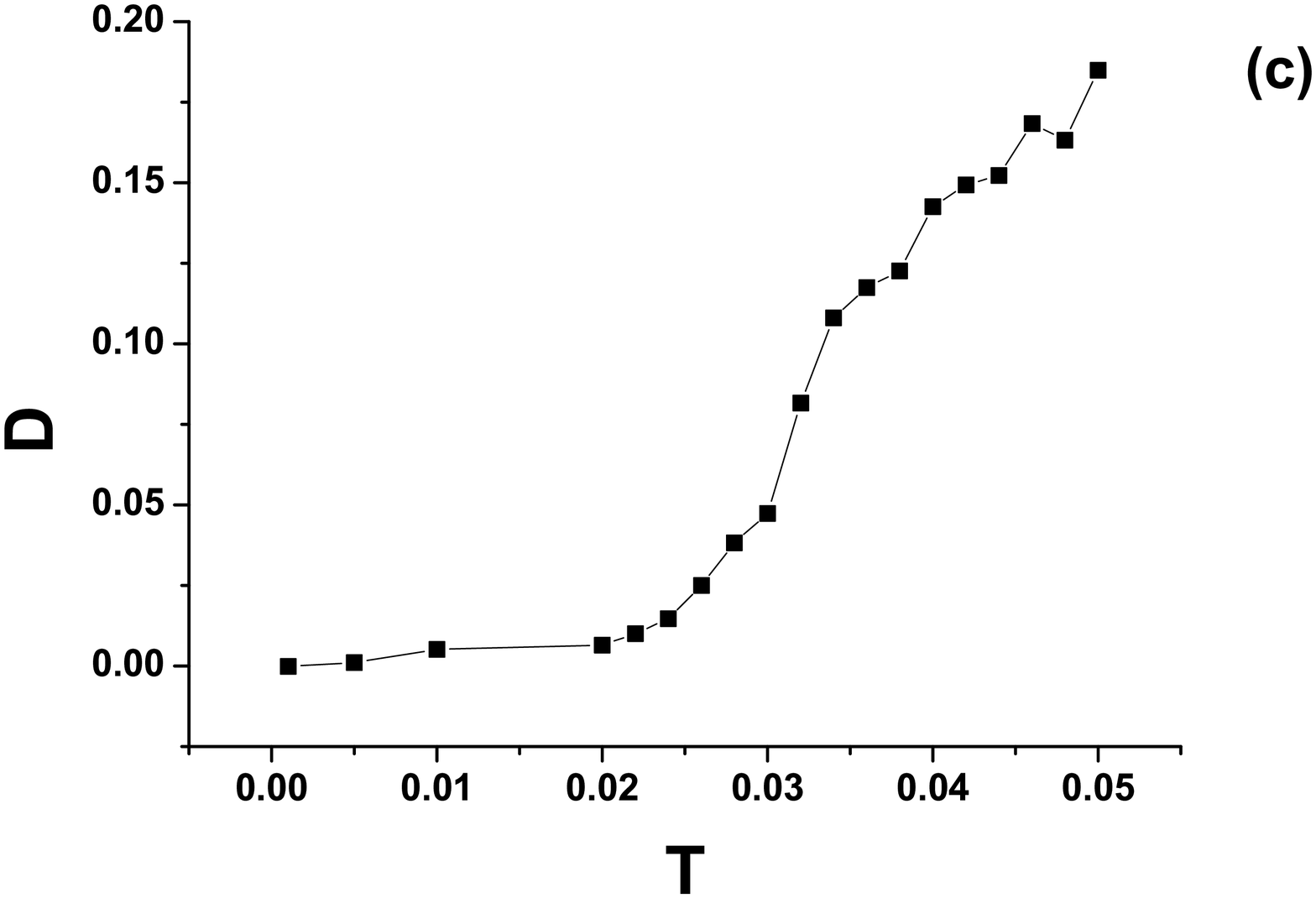}%

\caption{\label{r09} (a) Pressure of the oblique system along the
isochor $\rho=0.9$; (b) isochoric heat capacity along the same
isochor. The inset shows the internal energy along the same
isochor. (c) Diffusion coefficient of the same system.}
\end{figure}

We simulate the tilted system at constant density $\rho=0.9$ and
temperature from $T_{min}=0.005$ up to $T=0.05$ and at constant
pressure $P=3.0$ and temperature from $T=0.002$ up to $T=0.005$.
The tilt angle is set to $cos(\gamma)=0.415$. Fig. ~\ref{r09}
shows the pressure, isochoric heat capacity and diffusion
coefficient at the isochor $\rho=0.9$. The heat capacity is
obtained by numerical differentiation of the internal energy. One
can see that there is a bend in the pressure and diffusion
coefficients at $T=0.0032$. The heat capacity demonstrates a peak
at the same temperature. Similar peak was observed by Malescio and
Pelicane in Ref. \cite{malpel}. At the same time we see that the
diffusion coefficient at the temperatures above the heat capacity
peak has liquid-like magnitudes.

\begin{figure}
\includegraphics[width=6cm,height=6cm]{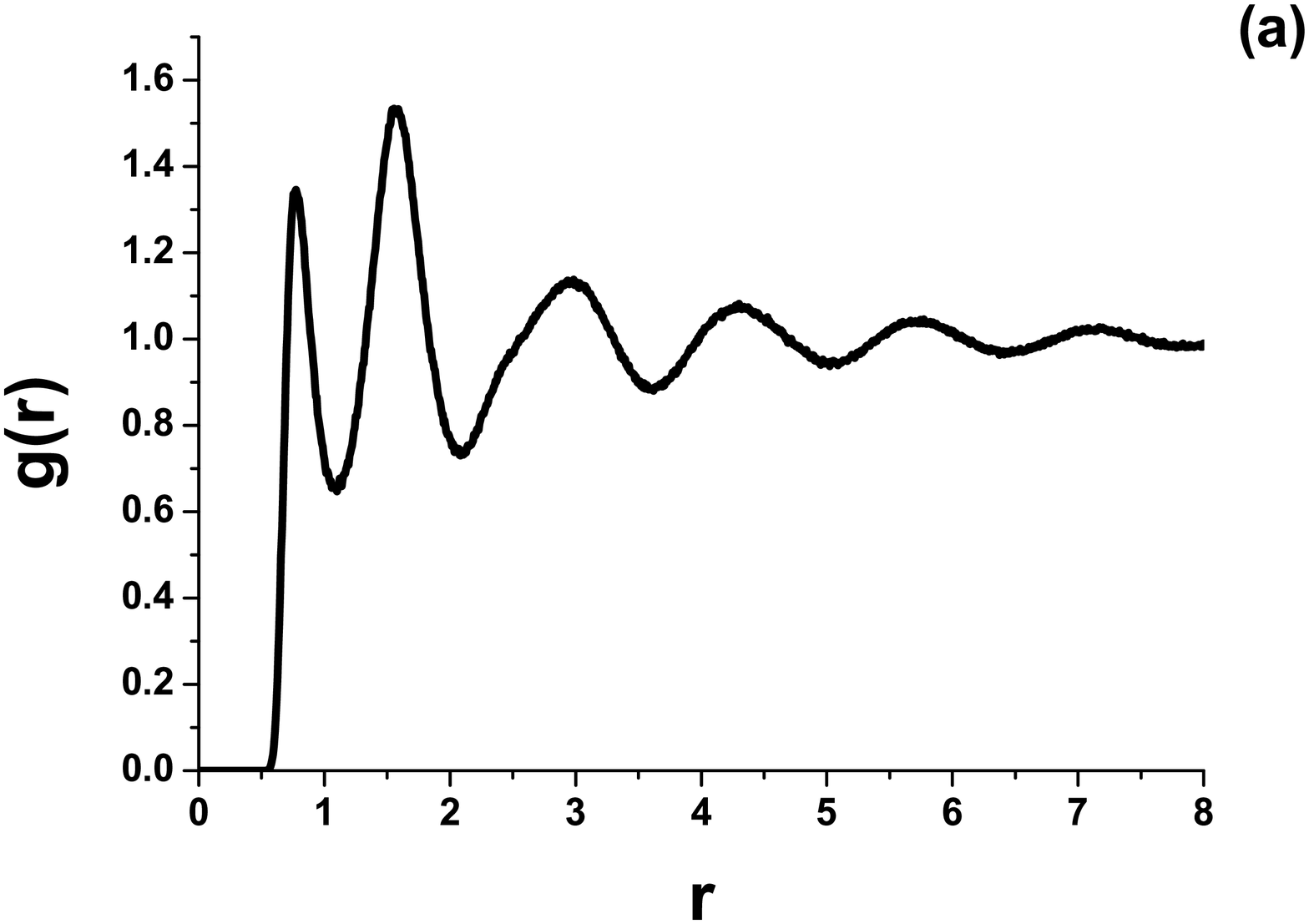}%

\includegraphics[width=6cm,height=4cm]{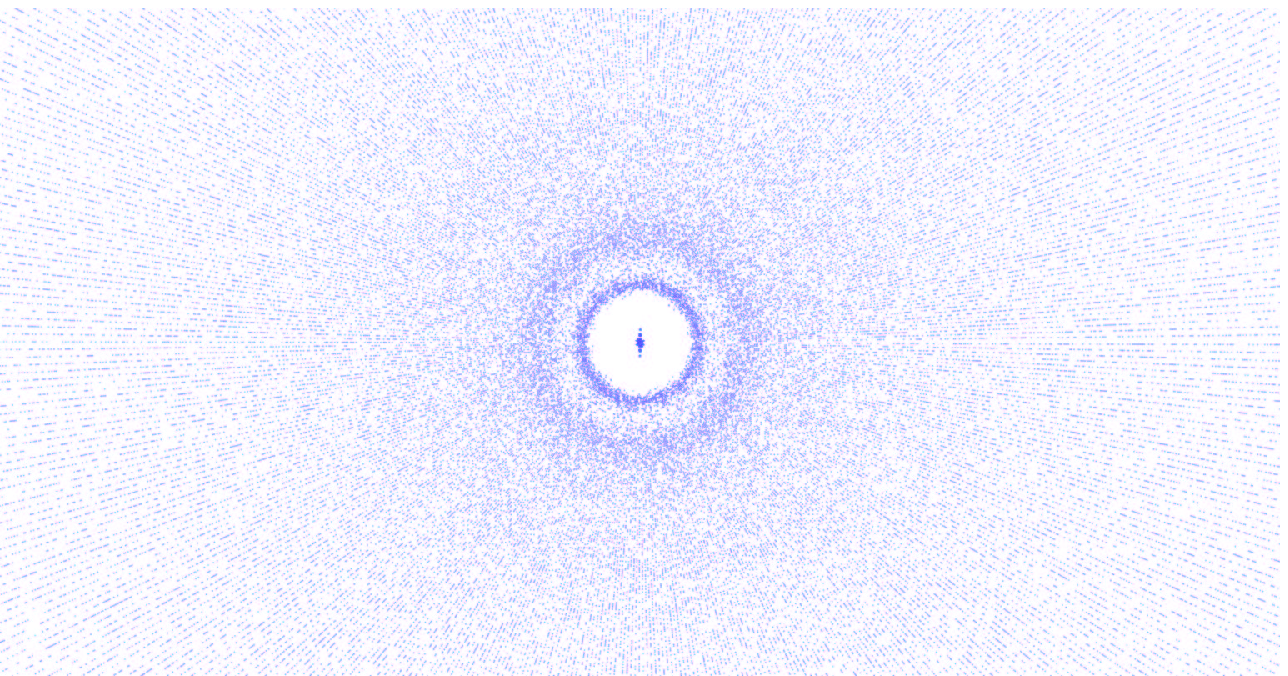}%

\caption{\label{r09-ht} (a) Radial distribution function at
$\rho=0.9$ and $T=0.04$. (b) Diffraction pattern at the same
point.}
\end{figure}

Fig. ~\ref{r09-ht} shows the rdf and diffraction pattern of the
system at $\rho=0.9$ and $T=0.04$, which is above the peak of the
heat capacity. Although rdf demonstrates several peaks it
corresponds to liquid phase. The same conclusion can be made from
the diffraction pattern. Combining it with large magnitude of the
diffusion coefficient we conclude that this phase is liquid.

Similar observations are made along the isobar $P=3.0$. The
corresponding plots are given in Supplementary materials.

Fig. \ref{p3} shows the equation of state and the isobaric heat
capacity. The heat capacity is obtained by numerical
differentiation of the enthalpy. One can see qualitatively the
same picture. The temperature dependence of density demonstrates a
bend at $T=0.032$ and the heat capacity has a peak at the same
temperature. The results on the heat capacities confirm that there
is a phase transition in the system which is in agreement with
Ref. \cite{malpel}.

\begin{figure}
\includegraphics[width=6cm,height=6cm]{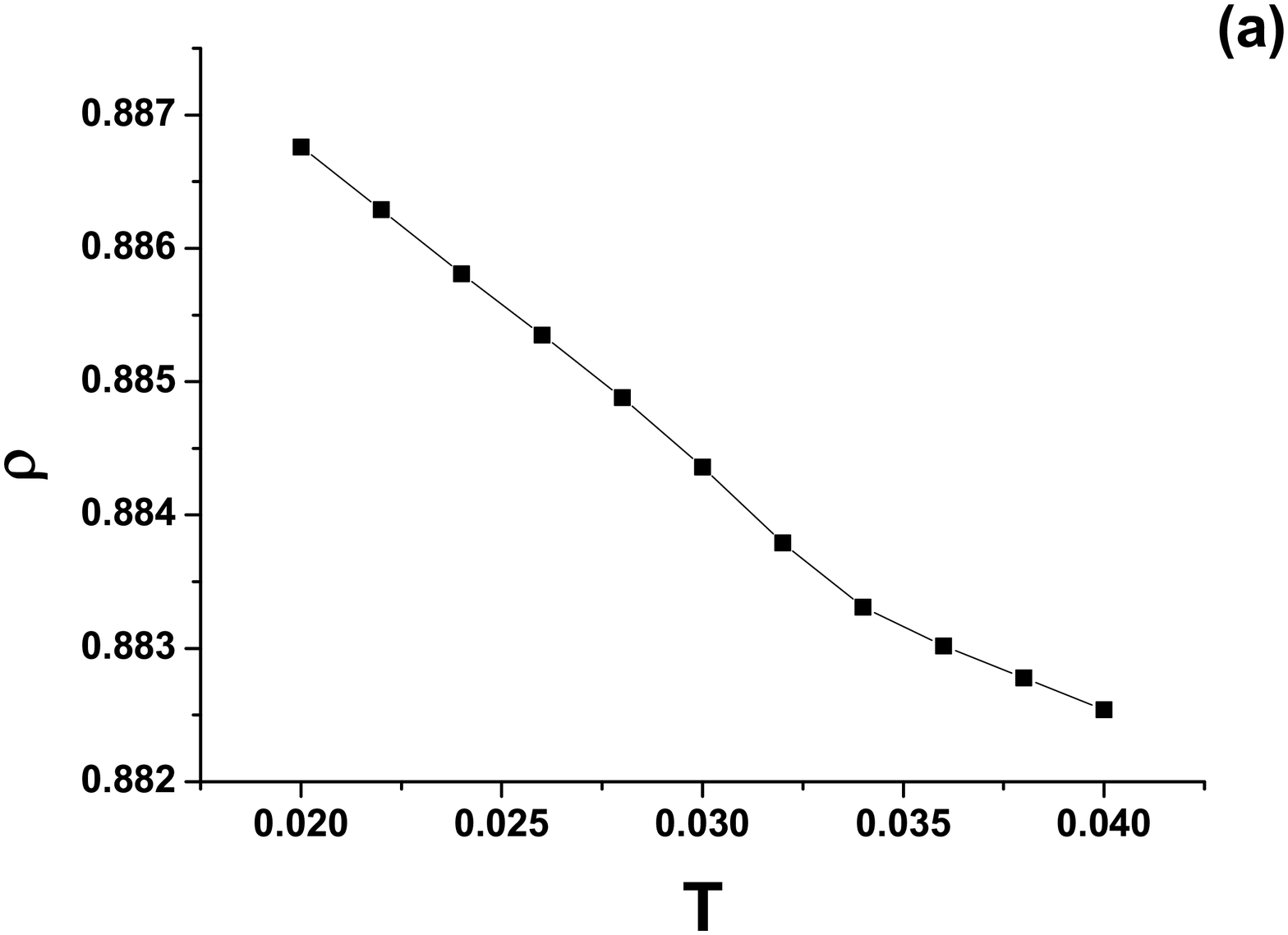}%

\includegraphics[width=6cm,height=6cm]{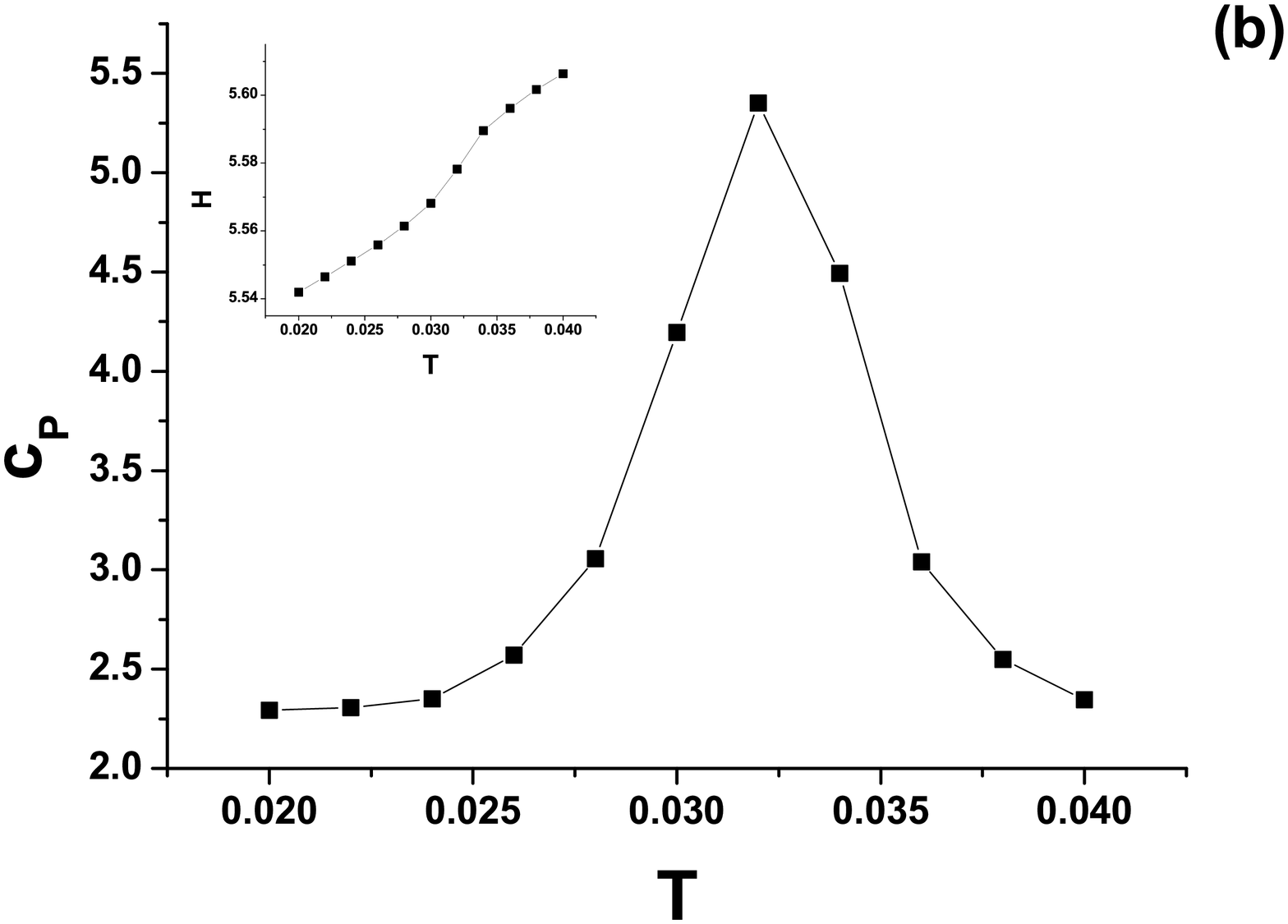}%

\caption{\label{p3} (a) Equation of state along the isobar $P=3.0$
(b) isobaric heat capacity along the same isobar. The inset shows
the enthalpy along the same isobar.}
\end{figure}

In conclusion, we find that stripe phase which is widely discussed
in the literature is indeed a crystalline phase made of tilted
triangles (oblique phase). The oblique phase was already observed
as a ground state structure of some core-softened systems
\cite{trusket2,genetic}, however it was not recognized that this
phase is the same with stripe phase which was reported by several
authors in core-softened systems at finite temperature. We find
the unit vectors of the oblique phase. When the temperature is
risen it transforms into correlated liquid. From the snapshots of
this liquid one can see that it contains some small stripes.
However, no long range order is observed in this liquid phase.

This work was carried out using computing resources of the federal
collective usage center "Complex for simulation and data
processing for mega-science facilities" at NRC "Kurchatov
Institute", http://ckp.nrcki.ru, and supercomputers at Joint
Supercomputer Center of the Russian Academy of Sciences (JSCC
RAS). The work was supported by the Russian Science Foundation
(Grant No 14-22-00093).

\end{document}